\documentclass[aps,twocolumn,prc,aps,superscriptaddress,amsmath,amssymb,floatfix,nofootinbib,superscriptaddress]{revtex4}

\usepackage{amsmath}
\usepackage{bm}
\usepackage{dcolumn}
\usepackage{subfigure}
\usepackage{amssymb}
\usepackage{xspace}
\usepackage{graphicx}

\usepackage{amssymb}
\usepackage{amsmath}
\usepackage{graphicx} 
\usepackage{epsfig}   
\usepackage{dcolumn}  
\usepackage{rotating} 

\usepackage{float}    
\floatstyle{boxed}

\usepackage{cases}
\usepackage[hidelinks=true]{hyperref}
\usepackage{breakurl}

\newcommand{\qsq}[0]{Q$^2$\xspace}

\begin{document}

\title{Separated kaon electroproduction cross section and the kaon form factor from 6 GeV JLab data}

\newcommand*{\CUA}{The Catholic University of America, Washington, DC 20064, USA}
\newcommand*{\JLAB}{Thomas Jefferson National Accelerator Facility, Newport News, VA 23606, USA}

\author{M.~Carmignotto} 
\affiliation{Catholic University of America, Washington, DC 20064}
\affiliation{Physics Division, TJNAF, Newport News, Virginia 23606}
\author{S.~Ali} 
\affiliation{Catholic University of America, Washington, DC 20064}
\author{K. Aniol}
\affiliation{California State University Los Angeles, Los Angeles, California
  90032}
\author{J. Arrington}
\affiliation{Physics Division, Argonne National Laboratory, Argonne, Illinois 60439}
\author{B. Barrett}
\affiliation{Saint Mary's University, Halifax, Nova Scotia, Canada}
\author{E.J. Beise}
\affiliation{University of Maryland, College Park, Maryland 20742}
\author{H.P. Blok}
\affiliation{Dept. of Physics, VU university, NL-1081 HV Amsterdam, The Netherlands}
\affiliation{NIKHEF, Postbus 41882, NL-1009 DB Amsterdam, The Netherlands}
\author{W. Boeglin}
\affiliation{Florida International University, Miami, Florida 33119}
\author{E.J. Brash}
\affiliation{University of Regina, Regina, Saskatchewan S4S 0A2, Canada}
\author{H. Breuer}
\affiliation{University of Maryland, College Park, Maryland 20742}
\author{C.C. Chang}
\affiliation{University of Maryland, College Park, Maryland 20742}
\author{M.E. Christy}
\affiliation{Hampton University, Hampton, Virginia 23668}
\author{A. Dittmann} 
\affiliation{University of Illinois, Champaign, Illinois 61801}
\author{R. Ent}
\affiliation{Physics Division, TJNAF, Newport News, Virginia 23606}
\author{H. Fenker}
\affiliation{Physics Division, TJNAF, Newport News, Virginia 23606}
\author{D. Gaskell}
\affiliation{Physics Division, TJNAF, Newport News, Virginia 23606}
\author{E. Gibson}
\affiliation{California State University, Sacramento, California 95819}
\author{R.J. Holt}
\affiliation{Physics Division, Argonne National Laboratory, Argonne, Illinois 60439}
\author{T.~Horn} 
\affiliation{Catholic University of America, Washington, DC 20064}
\affiliation{Physics Division, TJNAF, Newport News, Virginia 23606}
\author{G.M. Huber}
\affiliation{University of Regina, Regina, Saskatchewan S4S 0A2, Canada}
\author{S. Jin}
\affiliation{Kyungpook National University, Taegu, Korea}
\author{M.K. Jones}
\affiliation{College of William and Mary, Williamsburg, Virginia 23187}
\author{C.E. Keppel}
\affiliation{Hampton University, Hampton, Virginia 23668}
\affiliation{Physics Division, TJNAF, Newport News, Virginia 23606}
\author{W. Kim}
\affiliation{Kyungpook National University, Taegu, Korea}
\author{P.M. King}
\affiliation{University of Maryland, College Park, Maryland 20742}
\author{V. Kovaltchouk}
\affiliation{University of Regina, Regina, Saskatchewan S4S 0A2, Canada}
\author{J. Liu}
\affiliation{University of Maryland, College Park, Maryland 20742}
\author{G.J. Lolos}
\affiliation{University of Regina, Regina, Saskatchewan S4S 0A2, Canada}
\author{D.J. Mack}
\affiliation{Physics Division, TJNAF, Newport News, Virginia 23606}
\author{D.J. Margaziotis}
\affiliation{California State University Los Angeles, Los Angeles, California
  90032}
\author{P. Markowitz}
\affiliation{Florida International University, Miami, Florida 33119}
\author{A. Matsumura}
\affiliation{Tohoku University, Sendai, Japan}
\author{D. Meekins}
\affiliation{Physics Division, TJNAF, Newport News, Virginia 23606}
\author{T. Miyoshi}
\affiliation{Tohoku University, Sendai, Japan}
\author{H. Mkrtchyan}
\affiliation{Yerevan Physics Institute, 375036 Yerevan, Armenia}
\author{G. Niculescu}
\affiliation{James Madison University, Harrisonburg, Virginia 22807}
\author{I. Niculescu}
\affiliation{James Madison University, Harrisonburg, Virginia 22807}
\author{Y. Okayasu}
\affiliation{Tohoku University, Sendai, Japan}
\author{I. Pegg}
\affiliation{Catholic University of America, Washington, DC 20064}
\author{L. Pentchev}
\affiliation{College of William and Mary, Williamsburg, Virginia 23187}
\author{C. Perdrisat}
\affiliation{College of William and Mary, Williamsburg, Virginia 23187}
\author{D. Potterveld}
\affiliation{Physics Division, Argonne National Laboratory, Argonne, Illinois 60439}
\author{V. Punjabi}
\affiliation{Norfolk State University, Norfolk, Virginia}
\author{P. E. Reimer}
\affiliation{Physics Division, Argonne National Laboratory, Argonne, Illinois 60439}
\author{J. Reinhold}
\affiliation{Florida International University, Miami, Florida 33119}
\author{J. Roche}
\affiliation{Physics Division, TJNAF, Newport News, Virginia 23606}
\author{A. Sarty}
\affiliation{Saint Mary's University, Halifax, Nova Scotia, Canada}
\author{G.R. Smith}
\affiliation{Physics Division, TJNAF, Newport News, Virginia 23606}
\author{V. Tadevosyan}
\affiliation{Yerevan Physics Institute, 375036 Yerevan, Armenia}
\author{L.G. Tang}
\affiliation{Hampton University, Hampton, Virginia 23668}
\affiliation{Physics Division, TJNAF, Newport News, Virginia 23606}
\author{R.~Trotta} 
\affiliation{Catholic University of America, Washington, DC 20064}
\author{V. Tvaskis}
\affiliation{Faculteit Natuur- en Sterrenkunde, Vrije Universiteit, NL-1081 HV 
        Amsterdam, The Netherlands}
\author{A.~Vargas} 
\affiliation{Catholic University of America, Washington, DC 20064}
\author{S. Vidakovic}
\affiliation{University of Regina, Regina, Saskatchewan S4S 0A2, Canada}
\author{J. Volmer}
\affiliation{Faculteit Natuur- en Sterrenkunde, Vrije Universiteit, NL-1081 HV 
        Amsterdam, The Netherlands}
\affiliation{DESY, Hamburg, Germany}
\author{W. Vulcan}
\affiliation{Physics Division, TJNAF, Newport News, Virginia 23606}
\author{G. Warren}
\affiliation{Physics Division, TJNAF, Newport News, Virginia 23606}
\author{S.A. Wood}
\affiliation{Physics Division, TJNAF, Newport News, Virginia 23606}
\author{C. Xu}
\affiliation{University of Regina, Regina, Saskatchewan S4S 0A2, Canada}
\author{X. Zheng}
\affiliation{Physics Division, Argonne National Laboratory, Argonne, Illinois 60439}

\collaboration{and the Jefferson Lab FPI-2 and E93-018 Collaborations}
\noaffiliation

\date{\today}


\begin{abstract}
The $^{1}H$($e,e^\prime K^+$)$\Lambda$ reaction was studied as a function of the Mandelstam variable $-t$ using data from the E01-004 (FPI-2) and E93-018 experiments that were carried out in Hall~C at the 6 GeV Jefferson Lab. The cross section was fully separated into longitudinal and transverse components, and two interference terms at four-momentum transfers \qsq of 1.00, 1.36 and 2.07~GeV$^2$. The kaon form factor was extracted from the longitudinal cross section using the Regge model by Vanderhaeghen, Guidal, and Laget. The results establish the method, previously used successfully for pion analyses, for extracting the kaon form factor. Data from 12 GeV Jefferson Lab experiments are expected to have sufficient precision to distinguish between theoretical predictions, for example recent perturbative QCD calculations with modern parton distribution amplitudes. The leading-twist behavior for light mesons is predicted to set in for values of $Q^2$ between 5-10 GeV$^2$, which makes data in the few GeV regime particularly interesting. The $Q^2$ dependence at fixed $x$ and $-t$ of the longitudinal cross section we extracted seems consistent with the QCD factorization prediction within the experimental uncertainty.
\end{abstract}


\maketitle

The description of hadrons in terms of their constituents, the quarks and gluons, is a fundamental challenge in nuclear physics. Properties such as total charge and magnetic moments are well described in a constituent quark framework. However, charge and current distributions, which are more sensitive to the underlying dynamic processes in hadrons, are still not well described. The $^{1}H$($e,e^\prime K^+)\Lambda$ reaction provides the simplest system including strangeness, and is thus an effective experimental test of flavor degrees of freedom. 

The electromagnetic form factors of hadrons are directly connected to their internal structure. Measurements of the onset of the asymptotic, pointlike regime are an essential experimental verification of a key prediction of Quantum Chromodynamics (QCD)~\cite{Gao:2017mmp}. The form factors of light, two-quark hadronic systems like pions and kaons are of special importance as their asymptotic behavior is expected to set in earlier than that of three-quark systems. The relevance of pion and kaon form factors, for both experiment and theory, is evident in the literature~\cite{Frazer:1959zz,Farrar:1979aw,Nesterenko:1982gc,Amendolia:1986wj,Amendolia:1984nz,Bebek:1974ww,Bebek:1976qm,Bebek:1977pe,Ackermann:1977rp,Brauel:1979zk,Volmer:2000ek,Tadevosyan:2007yd,Blok:2008jy,Huber:2008id,Horn:2007ug,Horn:2006tm,Mecholsky:2017mpc,Gao:2017mmp}. A comprehensive review can be found in Ref.~\cite{Horn:2016rip}. 

A single form factor ($F_K$) determines the structure of the charged kaon, which is one of the simplest hadronic systems containing strangeness that is available for experimental study. Experimental values of the kaon form factor have been determined at low momentum transfers, $Q^2<$ 0.2GeV$^2$, by measurements of elastic scattering of high energy kaons off atomic electrons~\cite{Amendolia:1986ui}. Extending the reach of $F_K$ to higher values of $Q^2$ requires the use of meson electroproduction. As described in Refs.~\cite{Horn:2016rip,Blok:2008jy,Huber:2008id,Horn:2007ug,Horn:2006tm}, this method has been used successfully for extraction of the pion electroproduction cross section and form factor up to values of $Q^2$ of 3.91 GeV$^2$. 

The relative contribution of longitudinal and transverse terms to the meson cross section and their $-t$ and $Q^2$ dependencies are of great interest in evaluating the potential of probing the nucleon's transverse spatial structure through meson production. Recent calculations suggest that the leading-twist behavior for light mesons can be reached at values of $Q^2$=5-10 GeV$^2$~\cite{Horn:2016rip}. Experimental data to confirm these predictions are thus of great interest. In this paper, we present new, fully separated cross sections at values of \qsq = 1.00~GeV$^2$, 1.36 and 2.07~GeV$^2$, and analyze the current status of kaon electroproduction data in the few GeV regime. 

The goal of this paper is to extract the L/T separated kaon electroproduction cross section at different values of $-t$ from Jefferson Lab (JLab) E93-018~\cite{Mohring:2002tr} and FPI-2~\cite{Horn:2006tm} data and to apply the successful method from Refs.~\cite{Horn:2016rip,Blok:2008jy,Huber:2008id,Horn:2007ug,Horn:2006tm} to extract the kaon form factor. The separated longitudinal cross sections, scaled to constant $-t$ and $x$, will be used to analyze the $Q^2$ dependence of the data compared to QCD prediction.

The experiments were carried out in Hall C at the Thomas Jefferson National Accelerator Facility (Jefferson Lab), where two magnetic focusing spectrometers were used to detect charged final state particles. The data of these experiments were taken at different beam energies for values of $Q^2$ of 1.00~GeV$^2$ at a center of mass energy, $W$=1.81~GeV; 1.36 GeV, $W$=2.31~GeV; 2.07 GeV$^2$, $W$=2.31~GeV. Together with the control over systematic uncertainty offered by the focusing spectrometers, this allowed for an L-T separation of the kaon electroproduction cross section following the method described in Refs~\cite{Horn:2016rip,Blok:2008jy}.

In the E93-018 experiment, charged kaons were detected in the Short Orbit Spectrometer (SOS) while the scattered electrons were detected in the High Momentum Spectrometer (HMS). In the FPI-2 experiment, the detection was reversed to allow higher \qsq to be reached. Both spectrometers included two drift chambers for track reconstruction and scintillator arrays for triggering. To select electrons, a gas Cherenkov detector was used in combination with a lead-glass calorimeter in both experiments. Charged kaons were identified with an aerogel Cherenkov detector ($n$=1.034) in the SOS for E93-018 and a C$_4$F$_{10}$ gas Cherenkov detector in the HMS at 0.4 atm for FPI-2. Additional charged pion background was subtracted by fitting the background underneath the reconstructed missing mass. Any remaining contamination from real electron-proton coincidences was eliminated with a coincidence time cut of $\pm$ 0.75 ns. A detailed description of the SOS and the HMS can be found in Ref.~\cite{Blok:2008jy}.

\begin{figure}[!h]
\begin{center}
\includegraphics[width=0.47\textwidth]{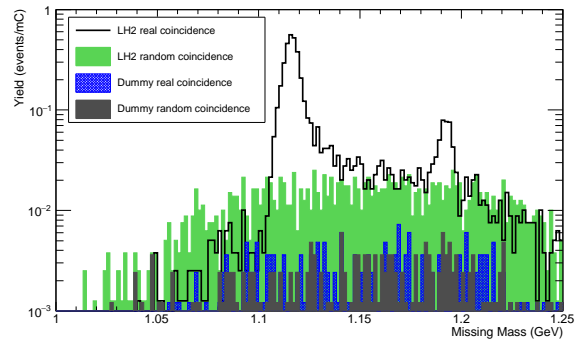}
\caption{\label{fig:background} (color online) Missing mass distribution of events within the real and random coincidence time window for the FPI-2 setting $Q^2$=2.07 GeV$^2$ and $-t$=0.39 (GeV/c)$^2$, integrated over the azimuthal angle. Also shown is the dummy target distribution.}
\end{center}
\end{figure}

The exclusive $\Lambda$ final state was selected with a cut on the reconstructed missing mass of the reaction. A representative coincidence spectrum is shown in Fig.~\ref{fig:background}. The random background is almost two orders of magnitude smaller than the real contribution under the $\Lambda$ peak. One can also clearly see the $\Sigma^0$ peak, which is well separated from the $\Lambda$ final state. Background from the aluminum target cell walls and random coincidences (together contributing 2-5\% of the yield after missing mass cuts) were subtracted from the experimental yields. The dependence of the cross section with different missing mass cuts applied, e.g. varying the upper limit between 1.12 and 1.18 GeV, was studied. The variation in the cross section is negligible below 1.15 GeV and less than 1\% for cuts above that value. To account for all backgrounds, a detailed background subtraction with functions representing the sum of the background and a peak centered at the $\Lambda$ mass was performed. The relevant electroproduction kinematic variables $Q^2$, $W$, and $-t$ were reconstructed from the measured spectrometer quantities. Experimental yields were calculated after correcting for inefficiencies, the dominant sources being tracking in the SOS (1.0\%) for E93-018 and acceptance in the HMS (0.7\%) for FPI-2. 

\begin{figure}[!h]
\begin{center}
\includegraphics[width=0.47\textwidth]{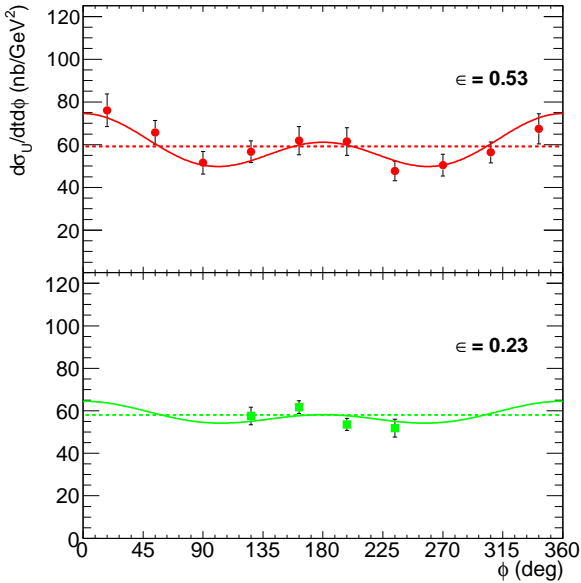}
\caption{\label{fig:xsec_fpi2_highQ2} (color online) Unseparated differential cross section extracted in different bins in $\phi$ for the two $\epsilon$ settings of the kinematics with \qsq=2.07~GeV$^2$ (W=2.31~GeV, $-t$=0.39~(GeV/c)$^2$) of the FPI-2 experiment. The dashed lines indicate the cross section averaged over all $\phi$ bins.}
\end{center}
\end{figure}

The unpolarized kaon electroproduction cross section can be written as the product of a virtual photon flux factor and a virtual photon cross section,
\begin{equation}
\frac{d^5\sigma}{d\Omega_edE_e^\prime dt d\phi}=\Gamma_\nu\frac{d^2\sigma}{d\Omega_K^*}\cdot J(t,\phi\rightarrow\Omega_K^*),
\label{eq:fivefold}
\end{equation}
\noindent where $\Gamma_\nu$ is a virtual photon flux factor, $\frac{d^2\sigma}{d\Omega_K^*}$ is the virtual photon production differential cross section at the kaon solid angle in the center of mass $d\Omega^*$, and $J$ is a Jacobian used to transform the production cross section in terms of the Mandelstam variable $-t$ and the angle $\phi$ between scattering and reaction planes. The information about the hadronic system is encoded in the virtual photon cross section. This can be expressed in terms of contributions from transversely and longitudinally polarized photons
\begin{equation}
\begin{aligned}
2\pi\frac{d^2\sigma_U}{dtd\phi}  = & \frac{d\sigma_T}{dt}+\epsilon\frac{d\sigma_L}{dt}+\sqrt{2\epsilon(1+\epsilon)}\frac{d\sigma_{LT}}{dt}\cos\phi \\
+ & \epsilon\frac{d\sigma_{TT}}{dt}\cos2\phi,
\label{eq:fourSigma}
\end{aligned}
\end{equation}
\noindent where $\epsilon$ is defined as the polarization of the virtual photon, $\epsilon=\left(1+\frac{2 |\mathbf{q}|^2}{Q^2} \tan^2\frac{\theta_{e}}{2} \right)^{-1}$. Here, $\mathbf{q}$ is the three-momentum of the transferred virtual photon and the electron scattering angle is denoted by $\theta_{e}$. The individual components in Eq.~\ref{eq:fourSigma} were determined from a simultaneous fit to the $\phi$ dependence of the measured cross sections at the different values of $\epsilon$. A representative example is shown in Fig.~\ref{fig:xsec_fpi2_highQ2}. 

The separated cross sections are determined at fixed values of $Q^2$, $W$, and $-t$, common for the different values of $\epsilon$. However, the acceptance covers a range in these quantities, so the measured yields represents an average over that range. To minimize errors from averaging, the experimental cross sections were calculated by comparing the experimental yield to a Monte Carlo simulation of the experiment. To account for variations of the cross section across the acceptance, the simulation uses a $^{1}H$($e, e^\prime, K^+$)$\Lambda$ model based on kaon electroproduction data~\cite{}. In addition, the Monte Carlo simulation includes a detailed description of the spectrometers, multiple scattering, ionization energy loss, kaon decay, and radiative processes. 

The uncertainties in the separated cross sections were estimated with both statistical and systematic sources. The E93-018 data were binned in two $-t$ bins with central values of 0.41 and 0.47 (GeV/c)$^2$ and six $\phi$ bins of width 60 degrees between 0 and 360 degrees. The statistical uncertainty in each bin ranges from 2.6-3.4\% at $-t$=0.41 (GeV/c)$^2$ and from 3.1-4.3\% at $-t$=0.41 (GeV/c)$^2$, summed over $\epsilon$. The FPI-2 data central $-t$ values are 0.27 (GeV/c)$^2$ and 0.39 (GeV/c)$^2$ for $Q^2$=1.36 and 2.07 GeV$^2$, respectively. Available statistics did not allow binning in $-t$. The data were binned in 10 $\phi$ bins at high $\epsilon$ and 4 $\phi$ bins at low $\epsilon$. The statistical uncertainties range from 2.2-5.2\% at $Q^2$=1.37 GeV$^2$ and from 2.1-4.4\% at $Q^2$=2.07 GeV$^2$ in these bins. 

Systematic uncertainties that are uncorrelated between the different $\epsilon$ points are amplified by a factor of 1/$\Delta \epsilon$ in the L-T separation. Correlated systematic uncertainties propagate directly into the separated cross sections. The uncorrelated systematic uncertainty is dominated by SOS tracking efficiency (1\%), radiative corrections (1\%), and kaon decay (1.2\%) for E93-018 and HMS and SOS acceptance (both 0.7\%), kaon decay (1\%) and absorption (0.5\%) for FPI-2 resulting in a total uncorrelated uncertainty of 2.3\% and 1.7\% respectively. The correlated systematic uncertainty is primarily due to kaon decay (3\%) for both experiments, and acceptance (2\%) and charge measurement (1.5\%) for E93-018, and radiative corrections (2\%), tracking efficiency and acceptance (all 1\%) for FPI-2 resulting in a total correlated uncertainty of 4.9\% and 4.5\% respectively. More details can be found in Refs.~\cite{Blok:2008jy,PHDMAPC}

To determine the kaon form factor, we compare the experimental results for $\sigma_L$ to a Regge model calculation by Vanderhaeghen, Guidal and Laget (VGL)~\cite{Vanderhaeghen:1997ts,Guidal:1997hy,Guidal:1997by,Guidal:1999qi}. Since most model parameters are fixed from photoproduction data, the cutoff of the Regge trajectories ($\Lambda_{K^+}$ and $\Lambda_{K^*}$) are the only free parameters. Both form factors are parameterized by a monopole form, $\left(1+Q^2/\Lambda_{K^+}^2\right)^{-1}$, where $\Lambda_K$ is taken as input to the model. The cutoff parameter $\Lambda_{K^*}^2$ is unknown and was considered to be equal to $\Lambda_K^2$. Varying $\Lambda_{K^*}^2$ between 0.1 and 1.5 GeV$^2$ changes $F_K$ by 0.7\%. Therefore, $F_K$ can be determined in a one-parameter fit from a comparison of the longitudinal cross section to the one predicted from the Regge model. 

\begin{figure}[!h]
\begin{center}
\includegraphics[width=0.47\textwidth]{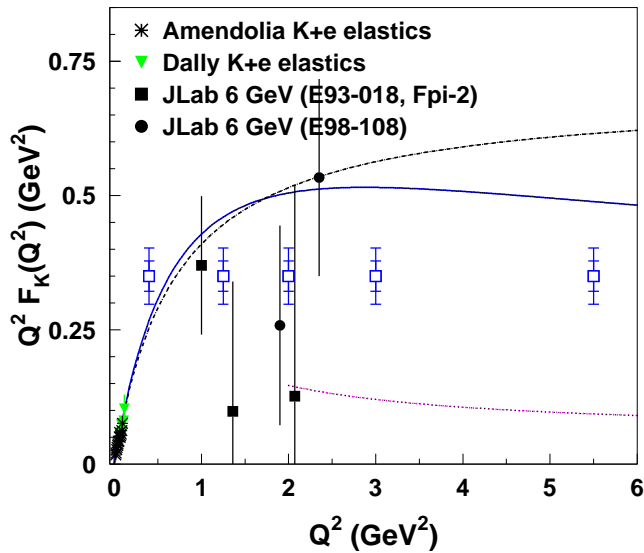}
\caption{\label{fig:Fk} (color online) Kaon form factor from E93-018 and FPI-2 (filled squares) compared to data from Refs.~\cite{Amendolia:1986ui,Dally:1980dj}. The error bars include the statistical and systematic uncertainty. Also shown are two kaon form factor points (filled circles) we extracted from the cross section data of Ref.~\cite{Coman:2009jk} using our form factor extraction method. The blue open squares indicate the projected reach and accuracy of data that are anticipated from JLab 12 GeV experiment E12-09-011~\cite{E12-09-011}. The two error estimates are based on different assumptions about the $-t$ and model dependence of the form factor extractions, with the larger uncertainty being more conservative. The dashed curve shows the monopole using a kaon charge radius of 0.56 fm obtained at low $Q^2$ in Ref.~\cite{Amendolia:1986ui}. The remaining curves are calculations with different kaon parton distribution amplitudes (PDAs). The solid blue line shows the kaon form factor obtained in Ref.~\cite{Gao:2017mmp} using the Dyson-Schwinger equation with rainbow ladder truncation. The curve is consistent within 15\% with that obtained using the leading-order (LO), leading-twist (LT) QCD prediction with a pion valence-quark PDA evaluated at a scale appropriate to the experiment. The dotted pink curve corresponds to a LO/LT calculation obtained with the conformal-limit PDA.}
\end{center}
\end{figure}

The VGL model gives a good description of the longitudinal cross section within the uncertainty. The value of $F_K$ was thus determined from a least squares fit of the Regge model prediction to the data.
The transverse cross section is described well in shape, but underpredicted in magnitude, similar to what was observed for the transverse pion cross section~\cite{Blok:2008jy,Horn:2007ug,Horn:2006tm,Horn:2012zza}. Possible explanations and further work on the size of the transverse cross section have been been presented in Refs.~\cite{Goloskokov:2011rd,Goloskokov:2009ia,Favart:2015umi}. 
%
Extensions to the VGL model to better describe the transverse cross section were developed in Refs.~\cite{Vrancx:2014pwa,DeCruz:2012bv,Kaskulov:2010kf}, which differ in the selection of the proton electromagnetic form factor. The model by Kaskulov and Mosel is only available for pion electroproduction cross sections. The longitudinal predictions of VGL and VR models are reasonably consistent over the $Q^2$-$W$ reach of our data though the difference between the two increases with increasing values of $Q^2$. 

The extraction of $F_K$ from $\sigma_L$ relies on dominance of the kaon exchange term. The kaon pole is farther from the physical region than the pion, which may raise doubts about the ability to extract $F_K$ from electroproduction data. To lend confidence to our method, we note two aspects. First, the pion form factor was extracted from pion electroproduction data at small $-t$ by carefully studying the model dependence of the analysis, not by direct extrapolation. We used the same method in our kaon form factor analysis. Second, comparative extractions of the pion form factor from low-$-t$ to large-$-t$ data, e.g. for $-t$=0.14-0.37 (GeV/c)$^2$ in Ref.~\cite{Horn:2016rip}, suggest only a modest model dependence. The largest $-t$ data lie at similar distances from the pole as our kaon data. We also note that a recent calculation~\cite{Qin:2017lcd} suggests that the kaon pole is dominant for $-t < 0.9$ (GeV/c)$^2$. Our data range from $-t$=0.5-0.7 (GeV/c)$^2$, and so fall into this regime.

In Fig.~\ref{fig:Fk}, our results are shown along with the results from Ref.~\cite{Amendolia:1986ui,Dally:1980dj}. We also show the results of extracting $F_K$ at $Q^2$=1.9 and 2.35 GeV$^2$ ($W$=2.14 GeV and 2.08 GeV, respectively) from the cross section data of Ref.~\cite{Coman:2009jk} using our method. We extracted the form factor at the $-t_{min}$ for each point (0.41 and 0.57 (GeV/c)$^2$). The total uncertainty includes both $\epsilon$-dependent and $-t$-correlated contributions. The higher $Q^2$ points from the FPI2 experiment are in addition statistics limited.  The model dependence was estimated using a comparative study at different values of $-t$ and was found to be on the order of about 0.1 on the form factor value.
The two data sets are internally consistent, lending confidence to our method for extracting the kaon form factor from longitudinal cross section data.

It has been pointed out that the choice of suitable Parton Distribution Amplitudes (PDAs) plays a crucial role in perturbative QCD calculations. We show our results along with a calculation using the continuum bound-state method, the Dyson-Schwinger equation rainbow-ladder truncation, of Ref.~\cite{Gao:2017mmp}. The calculation uses a leading-order (LO), leading-twist (LT) perturbative QCD calculation using the conformal-limit PDA. To distinguish between the theoretical curve and the monopole, a 10-15\% measurement at $Q^2$=4-5 GeV$^2$ would be needed. This could be achieved with JLab 12 GeV experiment E12-09-011~\cite{E12-09-011}. The projected uncertainties are shown. 

\begin{figure}[t!]
\begin{center}
\includegraphics[width=0.47\textwidth]{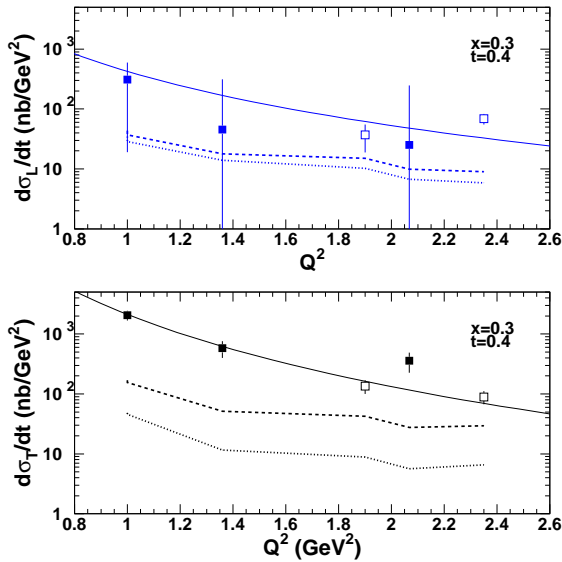}
\caption{\label{fig:xsec_sep} (color online) The Q$^2$-dependence of the separated cross sections at fixed values of $-t$=0.4 (GeV/c)$^2$ and $x_B$=0.3. The filled symbols are the scaled cross sections from E93-018 and FPI-2 and the open symbols from E98-108 data. The error bars denote the statistical and systematic uncertainties combined in quadrature. The solid curves show a fit of the form $Q^{-n}$, for $\sigma_L$ and $Q^{-m}$ for $\sigma_T$, where $n$=5.69$ \pm$ 0.52 and $m$=8.47 $\pm$ 0.46. Calculations using the VR (dashed) and the VGL (dotted) models are also shown.}
\end{center}
\end{figure}

One of the most stringent experimental tests of leading-twist behavior in meson production is the $Q^2$ dependence of the longitudinal meson cross section. In the regime where the leading-twist formalism is applicable, $\sigma_L$ is predicted to scale as Q$^{-6}$, the transverse cross section is expected to scale as $\sigma_T$ $\sim$ Q$^{-8}$ and $\sigma_L >> \sigma_T$. Fig.~\ref{fig:xsec_sep} shows the separated kaon cross sections from our analysis and points from Ref.~\cite{Coman:2009jk}. Here, we scaled all data to fixed values of $-t$=0.4 (GeV/c)$^2$ and $x$=0.3 using the separated Regge model by Vrancx and Ryckebusch (VR)~\cite{Vrancx:2014pwa} cross section predictions. The VR model extends the VGL Regge model by adding a hadronic model, which incorporates DIS light meson electroproduction at the amplitude level. 
The VR model transverse predictions are higher in magnitude by about a factor of four compared to those of the VGL model for all kinematic settings. At values of $Q^2 >$ 2.0 GeV$^2$ for $-t \sim x$, the VR and VGL model descriptions underpredict the data by a factor of four (five) compared to values of $-t >x$.

%
We have fitted $\sigma_L$ and $\sigma_T$ at each value of $x_B$ to the forms $\sigma_L \sim Q^{-n}$ and $\sigma_T \sim  Q^{-m}$, where $n$ and $m$ are free parameters. The experimental fit values are $n$=5.69 $\pm$ 0.52 (probability=98\%) and $m$=8.47 $\pm$ 0.46 (probability=97\%). The $\chi^2$ values for constant $n$=-6 and $m$=-8 are 1.68 and 2.10, respectively. The fit values for $\sigma_L$ appear to be consistent with the hard scattering prediction, however, the uncertainty on the data is large. In fact, fitting the $Q^2$-scaling prediction, $\sigma_L \sim Q^{-6}$, also results in a reasonable description of the data. While the scaling laws appear to be reasonably consistent with the $Q^2$-dependence of the $\sigma_L$ data, the $Q^2$-dependence of the $\sigma_T$ data is less well described. The $Q^2$-dependence of $\sigma_T$ does, however, provide less conclusive evidence for having reached the hard scattering regime as the factorization theorem was proven rigorously only for longitudinal photons. 

The contribution of transverse photons seems to be dominant over that of longitudinal ones for values of $Q^2$ up to 2 GeV$^2$. The ratio of transverse to longitudinal cross sections decreases from about ten at $Q^2$=1.0 to four at $Q^2 \sim$ 2.0 GeV$^2$. At $Q^2$=2.35 GeV$^2$, longitudinal and transverse cross sections are equal in magnitude to 10\%, within the uncertainty. If $\sigma_T$ can be confirmed to be dominant over a larger $Q^2$ range accessible with experiments at 12 GeV JLab, e.g. E12-09-011~\cite{E12-09-011}, this would allow one to probe transversity Generalized Parton Distributions as discussed in Refs.~\cite{Goloskokov:2011rd,Goloskokov:2009ia,Favart:2015umi,Goldstein:2012az,Ahmad:2008hp}.

In summary, we have determined separated kaon electroproduction cross sections at values of $Q^2$=1.00, 1.36 and 2.07 GeV$^2$ at $W$=1.81, and 2.31 GeV. The charged kaon form factor was extracted from the separated longitudinal cross section using the VGL Regge model. The results establish the method, previously used successfully for pion analyses, for extracting the kaon form factor. Planned kaon form factor data are expected to have sufficient precision to distinguish between the monopole and recent perturbative QCD calculations with modern parton distribution amplitudes. The leading-twist behavior for light mesons is predicted to set in for values of $Q^2$ between 5-10 GeV$^2$, which makes data in the few GeV regime particularly interesting. The $Q^2$ dependence at fixed $x$ and $-t$ of the longitudinal cross section we extracted appears to be consistent with the QCD factorization prediction within the experimental uncertainty.

\vspace{0.6in}
\centerline{ACKNOWLEDGMENTS} 
This work was supported in part by NSF grants PHY1714133, PHY1306227 and PHY1306418. This material is based upon work supported by the U.S. Department of Energy, Office of Science, Office of Nuclear Physics under contract DE-AC05-06OR23177. We acknowledge additional research grants from U.S. Department of Energy, ANL DOE grant: DE-AC02-06CH11357, the Natural Sciences and Engineering Research Council of Canada (NSERC), NATO and FOM (Netherlands).


\bibliography{thbibtanja}

\begin{thebibliography}{39}
\expandafter\ifx\csname natexlab\endcsname\relax\def\natexlab#1{#1}\fi
\expandafter\ifx\csname bibnamefont\endcsname\relax
  \def\bibnamefont#1{#1}\fi
\expandafter\ifx\csname bibfnamefont\endcsname\relax
  \def\bibfnamefont#1{#1}\fi
\expandafter\ifx\csname citenamefont\endcsname\relax
  \def\citenamefont#1{#1}\fi
\expandafter\ifx\csname url\endcsname\relax
  \def\url#1{\texttt{#1}}\fi
\expandafter\ifx\csname urlprefix\endcsname\relax\def\urlprefix{URL }\fi
\providecommand{\bibinfo}[2]{#2}
\providecommand{\eprint}[2][]{\url{#2}}

\bibitem[{\citenamefont{Gao et~al.}(2017)\citenamefont{Gao, Chang, Liu,
  Roberts, and Tandy}}]{Gao:2017mmp}
\bibinfo{author}{\bibfnamefont{F.}~\bibnamefont{Gao}},
  \bibinfo{author}{\bibfnamefont{L.}~\bibnamefont{Chang}},
  \bibinfo{author}{\bibfnamefont{Y.-X.} \bibnamefont{Liu}},
  \bibinfo{author}{\bibfnamefont{C.~D.} \bibnamefont{Roberts}},
  \bibnamefont{and} \bibinfo{author}{\bibfnamefont{P.~C.} \bibnamefont{Tandy}},
  \bibinfo{journal}{Phys. Rev.} \textbf{\bibinfo{volume}{D96}},
  \bibinfo{pages}{034024} (\bibinfo{year}{2017}), \eprint{1703.04875}.

\bibitem[{\citenamefont{Frazer}(1959)}]{Frazer:1959zz}
\bibinfo{author}{\bibfnamefont{W.~R.} \bibnamefont{Frazer}},
  \bibinfo{journal}{Phys. Rev.} \textbf{\bibinfo{volume}{115}},
  \bibinfo{pages}{1763} (\bibinfo{year}{1959}).

\bibitem[{\citenamefont{Farrar and Jackson}(1979)}]{Farrar:1979aw}
\bibinfo{author}{\bibfnamefont{G.~R.} \bibnamefont{Farrar}} \bibnamefont{and}
  \bibinfo{author}{\bibfnamefont{D.~R.} \bibnamefont{Jackson}},
  \bibinfo{journal}{Phys. Rev. Lett.} \textbf{\bibinfo{volume}{43}},
  \bibinfo{pages}{246} (\bibinfo{year}{1979}).

\bibitem[{\citenamefont{Nesterenko and Radyushkin}(1982)}]{Nesterenko:1982gc}
\bibinfo{author}{\bibfnamefont{V.~A.} \bibnamefont{Nesterenko}}
  \bibnamefont{and} \bibinfo{author}{\bibfnamefont{A.~V.}
  \bibnamefont{Radyushkin}}, \bibinfo{journal}{Phys. Lett.}
  \textbf{\bibinfo{volume}{115B}}, \bibinfo{pages}{410} (\bibinfo{year}{1982}).

\bibitem[{\citenamefont{Amendolia
  et~al.}(1986{\natexlab{a}})}]{Amendolia:1986wj}
\bibinfo{author}{\bibfnamefont{S.~R.} \bibnamefont{Amendolia}}
  \bibnamefont{et~al.} (\bibinfo{collaboration}{NA7}), \bibinfo{journal}{Nucl.
  Phys.} \textbf{\bibinfo{volume}{B277}}, \bibinfo{pages}{168}
  (\bibinfo{year}{1986}{\natexlab{a}}).

\bibitem[{\citenamefont{Amendolia et~al.}(1984)}]{Amendolia:1984nz}
\bibinfo{author}{\bibfnamefont{S.~R.} \bibnamefont{Amendolia}}
  \bibnamefont{et~al.}, \bibinfo{journal}{Phys. Lett.}
  \textbf{\bibinfo{volume}{B146}}, \bibinfo{pages}{116} (\bibinfo{year}{1984}).

\bibitem[{\citenamefont{Bebek et~al.}(1976{\natexlab{a}})\citenamefont{Bebek,
  Brown, Herzlinger, Holmes, Lichtenstein, Pipkin, Raither, and
  Sisterson}}]{Bebek:1974ww}
\bibinfo{author}{\bibfnamefont{C.~J.} \bibnamefont{Bebek}},
  \bibinfo{author}{\bibfnamefont{C.~N.} \bibnamefont{Brown}},
  \bibinfo{author}{\bibfnamefont{M.}~\bibnamefont{Herzlinger}},
  \bibinfo{author}{\bibfnamefont{S.~D.} \bibnamefont{Holmes}},
  \bibinfo{author}{\bibfnamefont{C.~A.} \bibnamefont{Lichtenstein}},
  \bibinfo{author}{\bibfnamefont{F.~M.} \bibnamefont{Pipkin}},
  \bibinfo{author}{\bibfnamefont{S.}~\bibnamefont{Raither}}, \bibnamefont{and}
  \bibinfo{author}{\bibfnamefont{L.~K.} \bibnamefont{Sisterson}},
  \bibinfo{journal}{Phys. Rev.} \textbf{\bibinfo{volume}{D13}},
  \bibinfo{pages}{25} (\bibinfo{year}{1976}{\natexlab{a}}).

\bibitem[{\citenamefont{Bebek et~al.}(1976{\natexlab{b}})}]{Bebek:1976qm}
\bibinfo{author}{\bibfnamefont{C.~J.} \bibnamefont{Bebek}}
  \bibnamefont{et~al.}, \bibinfo{journal}{Phys. Rev. Lett.}
  \textbf{\bibinfo{volume}{37}}, \bibinfo{pages}{1326}
  (\bibinfo{year}{1976}{\natexlab{b}}).

\bibitem[{\citenamefont{Bebek et~al.}(1978)}]{Bebek:1977pe}
\bibinfo{author}{\bibfnamefont{C.~J.} \bibnamefont{Bebek}}
  \bibnamefont{et~al.}, \bibinfo{journal}{Phys. Rev.}
  \textbf{\bibinfo{volume}{D17}}, \bibinfo{pages}{1693} (\bibinfo{year}{1978}).

\bibitem[{\citenamefont{Ackermann et~al.}(1978)\citenamefont{Ackermann,
  Azemoon, Gabriel, Mertiens, Reich, Specht, Janata, and
  Schmidt}}]{Ackermann:1977rp}
\bibinfo{author}{\bibfnamefont{H.}~\bibnamefont{Ackermann}},
  \bibinfo{author}{\bibfnamefont{T.}~\bibnamefont{Azemoon}},
  \bibinfo{author}{\bibfnamefont{W.}~\bibnamefont{Gabriel}},
  \bibinfo{author}{\bibfnamefont{H.~D.} \bibnamefont{Mertiens}},
  \bibinfo{author}{\bibfnamefont{H.~D.} \bibnamefont{Reich}},
  \bibinfo{author}{\bibfnamefont{G.}~\bibnamefont{Specht}},
  \bibinfo{author}{\bibfnamefont{F.}~\bibnamefont{Janata}}, \bibnamefont{and}
  \bibinfo{author}{\bibfnamefont{D.}~\bibnamefont{Schmidt}},
  \bibinfo{journal}{Nucl. Phys.} \textbf{\bibinfo{volume}{B137}},
  \bibinfo{pages}{294} (\bibinfo{year}{1978}).

\bibitem[{\citenamefont{Brauel et~al.}(1979)\citenamefont{Brauel, Canzler,
  Cords, Felst, Grindhammer, Helm, Kollmann, Krehbiel, and
  Schadlich}}]{Brauel:1979zk}
\bibinfo{author}{\bibfnamefont{P.}~\bibnamefont{Brauel}},
  \bibinfo{author}{\bibfnamefont{T.}~\bibnamefont{Canzler}},
  \bibinfo{author}{\bibfnamefont{D.}~\bibnamefont{Cords}},
  \bibinfo{author}{\bibfnamefont{R.}~\bibnamefont{Felst}},
  \bibinfo{author}{\bibfnamefont{G.}~\bibnamefont{Grindhammer}},
  \bibinfo{author}{\bibfnamefont{M.}~\bibnamefont{Helm}},
  \bibinfo{author}{\bibfnamefont{W.~D.} \bibnamefont{Kollmann}},
  \bibinfo{author}{\bibfnamefont{H.}~\bibnamefont{Krehbiel}}, \bibnamefont{and}
  \bibinfo{author}{\bibfnamefont{M.}~\bibnamefont{Schadlich}},
  \bibinfo{journal}{Z. Phys.} \textbf{\bibinfo{volume}{C3}},
  \bibinfo{pages}{101} (\bibinfo{year}{1979}).

\bibitem[{\citenamefont{Volmer et~al.}(2001)}]{Volmer:2000ek}
\bibinfo{author}{\bibfnamefont{J.}~\bibnamefont{Volmer}} \bibnamefont{et~al.}
  (\bibinfo{collaboration}{Jefferson Lab F(pi)}), \bibinfo{journal}{Phys. Rev.
  Lett.} \textbf{\bibinfo{volume}{86}}, \bibinfo{pages}{1713}
  (\bibinfo{year}{2001}), \eprint{nucl-ex/0010009}.

\bibitem[{\citenamefont{Tadevosyan et~al.}(2007)}]{Tadevosyan:2007yd}
\bibinfo{author}{\bibfnamefont{V.}~\bibnamefont{Tadevosyan}}
  \bibnamefont{et~al.} (\bibinfo{collaboration}{Jefferson Lab F(pi)}),
  \bibinfo{journal}{Phys. Rev.} \textbf{\bibinfo{volume}{C75}},
  \bibinfo{pages}{055205} (\bibinfo{year}{2007}), \eprint{nucl-ex/0607007}.

\bibitem[{\citenamefont{Blok et~al.}(2008)}]{Blok:2008jy}
\bibinfo{author}{\bibfnamefont{H.~P.} \bibnamefont{Blok}} \bibnamefont{et~al.}
  (\bibinfo{collaboration}{Jefferson Lab}), \bibinfo{journal}{Phys. Rev.}
  \textbf{\bibinfo{volume}{C78}}, \bibinfo{pages}{045202}
  (\bibinfo{year}{2008}), \eprint{0809.3161}.

\bibitem[{\citenamefont{Huber et~al.}(2008)}]{Huber:2008id}
\bibinfo{author}{\bibfnamefont{G.~M.} \bibnamefont{Huber}} \bibnamefont{et~al.}
  (\bibinfo{collaboration}{Jefferson Lab}), \bibinfo{journal}{Phys. Rev.}
  \textbf{\bibinfo{volume}{C78}}, \bibinfo{pages}{045203}
  (\bibinfo{year}{2008}), \eprint{0809.3052}.

\bibitem[{\citenamefont{Horn et~al.}(2008)}]{Horn:2007ug}
\bibinfo{author}{\bibfnamefont{T.}~\bibnamefont{Horn}} \bibnamefont{et~al.},
  \bibinfo{journal}{Phys. Rev.} \textbf{\bibinfo{volume}{C78}},
  \bibinfo{pages}{058201} (\bibinfo{year}{2008}), \eprint{0707.1794}.

\bibitem[{\citenamefont{Horn et~al.}(2006)}]{Horn:2006tm}
\bibinfo{author}{\bibfnamefont{T.}~\bibnamefont{Horn}} \bibnamefont{et~al.}
  (\bibinfo{collaboration}{Jefferson Lab F(pi)-2}), \bibinfo{journal}{Phys.
  Rev. Lett.} \textbf{\bibinfo{volume}{97}}, \bibinfo{pages}{192001}
  (\bibinfo{year}{2006}), \eprint{nucl-ex/0607005}.

\bibitem[{\citenamefont{Mecholsky et~al.}(2017)\citenamefont{Mecholsky,
  Meija-Ott, Carmignotto, Horn, Miller, and Pegg}}]{Mecholsky:2017mpc}
\bibinfo{author}{\bibfnamefont{N.~A.} \bibnamefont{Mecholsky}},
  \bibinfo{author}{\bibfnamefont{J.}~\bibnamefont{Meija-Ott}},
  \bibinfo{author}{\bibfnamefont{M.}~\bibnamefont{Carmignotto}},
  \bibinfo{author}{\bibfnamefont{T.}~\bibnamefont{Horn}},
  \bibinfo{author}{\bibfnamefont{G.~A.} \bibnamefont{Miller}},
  \bibnamefont{and} \bibinfo{author}{\bibfnamefont{I.~L.} \bibnamefont{Pegg}},
  \bibinfo{journal}{Phys. Rev.} \textbf{\bibinfo{volume}{C96}},
  \bibinfo{pages}{065207} (\bibinfo{year}{2017}), \eprint{1709.02853}.

\bibitem[{\citenamefont{Horn and Roberts}(2016)}]{Horn:2016rip}
\bibinfo{author}{\bibfnamefont{T.}~\bibnamefont{Horn}} \bibnamefont{and}
  \bibinfo{author}{\bibfnamefont{C.~D.} \bibnamefont{Roberts}},
  \bibinfo{journal}{J. Phys.} \textbf{\bibinfo{volume}{G43}},
  \bibinfo{pages}{073001} (\bibinfo{year}{2016}), \eprint{1602.04016}.

\bibitem[{\citenamefont{Amendolia
  et~al.}(1986{\natexlab{b}})}]{Amendolia:1986ui}
\bibinfo{author}{\bibfnamefont{S.~R.} \bibnamefont{Amendolia}}
  \bibnamefont{et~al.}, \bibinfo{journal}{Phys. Lett.}
  \textbf{\bibinfo{volume}{B178}}, \bibinfo{pages}{435}
  (\bibinfo{year}{1986}{\natexlab{b}}).

\bibitem[{\citenamefont{Mohring et~al.}(2003)}]{Mohring:2002tr}
\bibinfo{author}{\bibfnamefont{R.~M.} \bibnamefont{Mohring}}
  \bibnamefont{et~al.} (\bibinfo{collaboration}{E93018}),
  \bibinfo{journal}{Phys. Rev.} \textbf{\bibinfo{volume}{C67}},
  \bibinfo{pages}{055205} (\bibinfo{year}{2003}), \eprint{nucl-ex/0211005}.

\bibitem[{\citenamefont{{Carmignotto M.}}(2017)}]{PHDMAPC}
\bibinfo{author}{\bibnamefont{{Carmignotto M.}}} (\bibinfo{year}{2017}),
  \bibinfo{note}{phD thesis Catholic University of America}.

\bibitem[{\citenamefont{Vanderhaeghen et~al.}(1998)\citenamefont{Vanderhaeghen,
  Guidal, and Laget}}]{Vanderhaeghen:1997ts}
\bibinfo{author}{\bibfnamefont{M.}~\bibnamefont{Vanderhaeghen}},
  \bibinfo{author}{\bibfnamefont{M.}~\bibnamefont{Guidal}}, \bibnamefont{and}
  \bibinfo{author}{\bibfnamefont{J.~M.} \bibnamefont{Laget}},
  \bibinfo{journal}{Phys. Rev.} \textbf{\bibinfo{volume}{C57}},
  \bibinfo{pages}{1454} (\bibinfo{year}{1998}).

\bibitem[{\citenamefont{Guidal et~al.}(1997{\natexlab{a}})\citenamefont{Guidal,
  Laget, and Vanderhaeghen}}]{Guidal:1997hy}
\bibinfo{author}{\bibfnamefont{M.}~\bibnamefont{Guidal}},
  \bibinfo{author}{\bibfnamefont{J.~M.} \bibnamefont{Laget}}, \bibnamefont{and}
  \bibinfo{author}{\bibfnamefont{M.}~\bibnamefont{Vanderhaeghen}},
  \bibinfo{journal}{Nucl. Phys.} \textbf{\bibinfo{volume}{A627}},
  \bibinfo{pages}{645} (\bibinfo{year}{1997}{\natexlab{a}}).

\bibitem[{\citenamefont{Guidal et~al.}(1997{\natexlab{b}})\citenamefont{Guidal,
  Laget, and Vanderhaeghen}}]{Guidal:1997by}
\bibinfo{author}{\bibfnamefont{M.}~\bibnamefont{Guidal}},
  \bibinfo{author}{\bibfnamefont{J.~M.} \bibnamefont{Laget}}, \bibnamefont{and}
  \bibinfo{author}{\bibfnamefont{M.}~\bibnamefont{Vanderhaeghen}},
  \bibinfo{journal}{Phys. Lett.} \textbf{\bibinfo{volume}{B400}},
  \bibinfo{pages}{6} (\bibinfo{year}{1997}{\natexlab{b}}).

\bibitem[{\citenamefont{Guidal et~al.}(2000)\citenamefont{Guidal, Laget, and
  Vanderhaeghen}}]{Guidal:1999qi}
\bibinfo{author}{\bibfnamefont{M.}~\bibnamefont{Guidal}},
  \bibinfo{author}{\bibfnamefont{J.~M.} \bibnamefont{Laget}}, \bibnamefont{and}
  \bibinfo{author}{\bibfnamefont{M.}~\bibnamefont{Vanderhaeghen}},
  \bibinfo{journal}{Phys. Rev.} \textbf{\bibinfo{volume}{C61}},
  \bibinfo{pages}{025204} (\bibinfo{year}{2000}), \eprint{hep-ph/9904511}.

\bibitem[{\citenamefont{Dally et~al.}(1980)}]{Dally:1980dj}
\bibinfo{author}{\bibfnamefont{E.~B.} \bibnamefont{Dally}}
  \bibnamefont{et~al.}, \bibinfo{journal}{Phys. Rev. Lett.}
  \textbf{\bibinfo{volume}{45}}, \bibinfo{pages}{232} (\bibinfo{year}{1980}).

\bibitem[{\citenamefont{Coman et~al.}(2010)}]{Coman:2009jk}
\bibinfo{author}{\bibfnamefont{M.}~\bibnamefont{Coman}} \bibnamefont{et~al.}
  (\bibinfo{collaboration}{Jefferson Lab Hall A}), \bibinfo{journal}{Phys.
  Rev.} \textbf{\bibinfo{volume}{C81}}, \bibinfo{pages}{052201}
  (\bibinfo{year}{2010}), \eprint{0911.3943}.

\bibitem[{\citenamefont{{Horn T., Huber G.M., Markowitz P., and
  others}}(2009)}]{E12-09-011}
\bibinfo{author}{\bibnamefont{{Horn T., Huber G.M., Markowitz P., and others}}}
  (\bibinfo{year}{2009}), \bibinfo{note}{approved Jefferson Lab 12 GeV
  Experiment}.

\bibitem[{\citenamefont{Horn}(2012)}]{Horn:2012zza}
\bibinfo{author}{\bibfnamefont{T.}~\bibnamefont{Horn}}, \bibinfo{journal}{Phys.
  Rev.} \textbf{\bibinfo{volume}{C85}}, \bibinfo{pages}{018202}
  (\bibinfo{year}{2012}).

\bibitem[{\citenamefont{Goloskokov and Kroll}(2011)}]{Goloskokov:2011rd}
\bibinfo{author}{\bibfnamefont{S.~V.} \bibnamefont{Goloskokov}}
  \bibnamefont{and} \bibinfo{author}{\bibfnamefont{P.}~\bibnamefont{Kroll}},
  \bibinfo{journal}{Eur. Phys. J.} \textbf{\bibinfo{volume}{A47}},
  \bibinfo{pages}{112} (\bibinfo{year}{2011}), \eprint{1106.4897}.

\bibitem[{\citenamefont{Goloskokov and Kroll}(2010)}]{Goloskokov:2009ia}
\bibinfo{author}{\bibfnamefont{S.~V.} \bibnamefont{Goloskokov}}
  \bibnamefont{and} \bibinfo{author}{\bibfnamefont{P.}~\bibnamefont{Kroll}},
  \bibinfo{journal}{Eur. Phys. J.} \textbf{\bibinfo{volume}{C65}},
  \bibinfo{pages}{137} (\bibinfo{year}{2010}), \eprint{0906.0460}.

\bibitem[{\citenamefont{Favart et~al.}(2016)\citenamefont{Favart, Guidal, Horn,
  and Kroll}}]{Favart:2015umi}
\bibinfo{author}{\bibfnamefont{L.}~\bibnamefont{Favart}},
  \bibinfo{author}{\bibfnamefont{M.}~\bibnamefont{Guidal}},
  \bibinfo{author}{\bibfnamefont{T.}~\bibnamefont{Horn}}, \bibnamefont{and}
  \bibinfo{author}{\bibfnamefont{P.}~\bibnamefont{Kroll}},
  \bibinfo{journal}{Eur. Phys. J.} \textbf{\bibinfo{volume}{A52}},
  \bibinfo{pages}{158} (\bibinfo{year}{2016}), \eprint{1511.04535}.

\bibitem[{\citenamefont{Vrancx et~al.}(2014)\citenamefont{Vrancx, Ryckebusch,
  and Nys}}]{Vrancx:2014pwa}
\bibinfo{author}{\bibfnamefont{T.}~\bibnamefont{Vrancx}},
  \bibinfo{author}{\bibfnamefont{J.}~\bibnamefont{Ryckebusch}},
  \bibnamefont{and} \bibinfo{author}{\bibfnamefont{J.}~\bibnamefont{Nys}},
  \bibinfo{journal}{Phys. Rev.} \textbf{\bibinfo{volume}{C89}},
  \bibinfo{pages}{065202} (\bibinfo{year}{2014}), \eprint{1404.4156}.

\bibitem[{\citenamefont{De~Cruz et~al.}(2012)\citenamefont{De~Cruz, Ryckebusch,
  Vrancx, and Vancraeyveld}}]{DeCruz:2012bv}
\bibinfo{author}{\bibfnamefont{L.}~\bibnamefont{De~Cruz}},
  \bibinfo{author}{\bibfnamefont{J.}~\bibnamefont{Ryckebusch}},
  \bibinfo{author}{\bibfnamefont{T.}~\bibnamefont{Vrancx}}, \bibnamefont{and}
  \bibinfo{author}{\bibfnamefont{P.}~\bibnamefont{Vancraeyveld}},
  \bibinfo{journal}{Phys. Rev.} \textbf{\bibinfo{volume}{C86}},
  \bibinfo{pages}{015212} (\bibinfo{year}{2012}), \eprint{1205.2195}.

\bibitem[{\citenamefont{Kaskulov and Mosel}(2010)}]{Kaskulov:2010kf}
\bibinfo{author}{\bibfnamefont{M.~M.} \bibnamefont{Kaskulov}} \bibnamefont{and}
  \bibinfo{author}{\bibfnamefont{U.}~\bibnamefont{Mosel}},
  \bibinfo{journal}{Phys. Rev.} \textbf{\bibinfo{volume}{C81}},
  \bibinfo{pages}{045202} (\bibinfo{year}{2010}), \eprint{1001.1952}.

\bibitem[{\citenamefont{Qin et~al.}(2017)\citenamefont{Qin, Chen, Mezrag, and
  Roberts}}]{Qin:2017lcd}
\bibinfo{author}{\bibfnamefont{S.-X.} \bibnamefont{Qin}},
  \bibinfo{author}{\bibfnamefont{C.}~\bibnamefont{Chen}},
  \bibinfo{author}{\bibfnamefont{C.}~\bibnamefont{Mezrag}}, \bibnamefont{and}
  \bibinfo{author}{\bibfnamefont{C.~D.} \bibnamefont{Roberts}}
  (\bibinfo{year}{2017}), \eprint{1702.06100}.

\bibitem[{\citenamefont{Goldstein et~al.}(2012)\citenamefont{Goldstein,
  Gonzalez~Hernandez, and Liuti}}]{Goldstein:2012az}
\bibinfo{author}{\bibfnamefont{G.~R.} \bibnamefont{Goldstein}},
  \bibinfo{author}{\bibfnamefont{J.~O.} \bibnamefont{Gonzalez~Hernandez}},
  \bibnamefont{and} \bibinfo{author}{\bibfnamefont{S.}~\bibnamefont{Liuti}},
  \bibinfo{journal}{J. Phys.} \textbf{\bibinfo{volume}{G39}},
  \bibinfo{pages}{115001} (\bibinfo{year}{2012}), \eprint{1201.6088}.

\bibitem[{\citenamefont{Ahmad et~al.}(2009)\citenamefont{Ahmad, Goldstein, and
  Liuti}}]{Ahmad:2008hp}
\bibinfo{author}{\bibfnamefont{S.}~\bibnamefont{Ahmad}},
  \bibinfo{author}{\bibfnamefont{G.~R.} \bibnamefont{Goldstein}},
  \bibnamefont{and} \bibinfo{author}{\bibfnamefont{S.}~\bibnamefont{Liuti}},
  \bibinfo{journal}{Phys. Rev.} \textbf{\bibinfo{volume}{D79}},
  \bibinfo{pages}{054014} (\bibinfo{year}{2009}), \eprint{0805.3568}.

\end{thebibliography}

\end{document}